\documentclass[lettersize,journal]{IEEEtran}
\usepackage{amsmath,amsfonts}
\usepackage{algorithmic}
\usepackage{algorithm}
\usepackage{array}
\usepackage[caption=false,font=normalsize,labelfont=sf,textfont=sf]{subfig}
\usepackage{textcomp}
\usepackage{stfloats}
\usepackage{url}
\usepackage{verbatim}
\usepackage{graphicx}
\usepackage{cite}
\usepackage{booktabs}
\begin{document}

\title{Multimodal LLM-based Query Paraphrasing for Video Search}

\author{Jiaxin Wu,~\IEEEmembership{Member,~IEEE}, Chong-Wah Ngo,~\IEEEmembership{Senior Member,~IEEE}, Wing-Kwong Chan, Sheng-Hua Zhong, Xiong-Yong Wei,~\IEEEmembership{Senior Member,~IEEE}, Qing Li,~\IEEEmembership{Fellow,~IEEE},
\thanks{
Jiaxin Wu is with School of Artificial Intelligence, Shenzhen University, China.}
\thanks{Jiaxin Wu, Xiong-Yong Wei, and Qing Li are with Department of Computing, The Hong Kong Polytechnic University, Hong Kong. 
 }
\thanks{Chong-Wah Ngo is with School of Computing and Information System, Singapore Management University, Singapore.}
\thanks{Wing-Kwong Chan is with Department of Computer Science, City University of Hong Kong, Hong Kong.}
\thanks{Sheng-Hua Zhong is with Department of Computer Science and Software Engineering, Shenzhen Univeristy, China.}
}

\markboth{Journal of \LaTeX\ Class Files,~Vol.~14, No.~8, August~2021}%
{Shell \MakeLowercase{\textit{et al.}}: A Sample Article Using IEEEtran.cls for IEEE Journals}


\maketitle

\begin{abstract}
Text-to-video retrieval answers user queries through searches based on concepts and embeddings. However, due to limitations in the size of the concept bank and the amount of training data, answering queries in the wild is not always effective because of the out-of-vocabulary problem. Furthermore, neither concept-based nor embedding-based search can perform reasoning to consolidate search results for complex queries that include logical and spatial constraints. To address these challenges, we leverage large language models (LLMs) to paraphrase queries using text-to-text (T2T), text-to-image (T2I), and image-to-text (I2T) transformations. These transformations rephrase abstract concepts into simpler terms to mitigate the out-of-vocabulary problem. Additionally, complex relationships within a query can be decomposed into simpler sub-queries, improving retrieval performance by effectively fusing the search results of these sub-queries. To mitigate the issue of LLM hallucination, this paper also proposes a novel consistency-based verification strategy to filter out factually incorrect paraphrased queries. Extensive experiments are conducted for ad-hoc video search and known-item search on the TRECVid datasets. We provide empirical insights into how traditionally difficult-to-answer queries can be effectively resolved through query paraphrasing.
\end{abstract}

\begin{IEEEkeywords}
Generative Queries, Query Paraphrasing, Out-of-vocabulary, Text-to-video Search
\end{IEEEkeywords}


\section{Introduction}
\IEEEPARstart{C}oncept-based \cite{Waseda2016,Jiang:semantic_gap,Lu:ICMR,Informedia2016} and embedding-based searches \cite{w2vvpp,tpami21-dual-encoding, shu_TMM2024MAC,zhangTextVideoRetrievalGLSCL2025} are two mainstream approaches for answering text-to-video search queries.
The concept-based approach maps a user query into tokens and measures similarity between the tokens and concepts extracted from videos. The embedding-based approach projects both the user query and video into a joint latent space, computing query-video similarity based on their embedding features.
Methods to combine them, such as the dual-task model \cite{dual_task} via dual-task learning, to understand a user query by both concepts and embedding features, are emerging. 

However, the effectiveness of existing methods is hindered by the size of the concept bank and the amount of training data. Directly answering an ad-hoc user query is not always effective due to the out-of-vocabulary (OOV) problem \cite{TRECVid2017, Trecvid2018}.
In other words, if no relevant concepts exist in the concept bank for a user query, or if the query is associated with only a small amount of training data, neither concept-based nor embedding-based search can achieve good retrieval performance. 
Several studies have addressed the OOV problem: Ueki et al. built large concept banks with over 55k concepts \cite{WasedaMeiseiSoftbank2019, wasedaAVS2020}. Ueki et al. proposed selecting concepts through ontology reasoning on WordNet \cite{Waseda2018}. Wu et al. generated pseudo captions for web videos, creating 7 million text-to-video pairs for pre-training concepts and embedding features \cite{improvedITV}. Although these methods have alleviated the OOV problem in video retrieval, they still inherently incur the OOV problem for an open-vocabulary query due to the size and variety of the concept banks used. Webly-labeled learning \cite{Informedia2018} attempts to solve the OOV problem by crawling related videos from the Internet to learn the concepts of a query. However, its effectiveness is limited by the amount of related data available on the Web and whether the relevant data can be retrieved.
Beyond the OOV problem, the retrieval performance of existing methods is also constrained by complex text-to-video search queries that involve:
logical constraints (e.g., "AND," "OR," "NOT"), and
spatial constraints (e.g., "in front of," "located in the lower corner of the scene"),
as evidenced by the TRECVid AVS evaluation \cite{Trecvid2019,trecvid2023}. There are very few studies addressing these complex queries, and those that do typically focus on a single type of complexity.
For example, BNL \cite{Understand_Negation_rumcc_MM} aims to interpret "NOT" in user queries for text-to-video retrieval by applying negation learning to positive and negative text-video pairs. It generates negative queries by adding "NOT" before verbs or after auxiliary verbs (if present) in a positive query.
In addition to the OOV and complex query problems, current text-to-video search methods also suffer from a representation robustness problem. Since text-to-video representation learning is trained in a one-to-one matching manner, retrieval results are highly sensitive to query formulation. Even slight variations in query wording can lead to significantly different retrieval results.
For example, in our experiment, the following two queries produced dramatically different retrieval results:
\textit{Find shots of an adult person running in a city street}.
\textit{Find shots of an adult running along a city street}.
Data augmentation has proven effective in enhancing the robustness of cross-modal representation learning \cite{LaCLIP,robustness_data_augumentation}. For instance, ChatGPT-QE has demonstrated that query expansion using ChatGPT can improve retrieval performance for ad-hoc video queries \cite{Waseda2023}.

To address robustness, OOV, and complex query problems in text-to-video retrieval, in this paper, we propose a novel LLM-based query paraphrasing framework to rephrase a user query by text-to-text (T2T), text-to-image (T2I), and image-to-text (I2T) transformations. The T2T transformation follows the effort of chatGPT-QE \cite{Waseda2023} to address the robustness problem in text-to-video matching by query augmentations. Specifically, we use an LLM to rewrite a given user query in multiple ways, generating query perturbations for text-to-video search. These rewritten queries may help mitigate the OOV problem if they replace OOV words with alternatives that coincidentally exist in the training data of the video search engine.
To further address the OOV issue, we introduce the T2I transformation, which leverages a text-to-image generation model to create various images using the user query as a prompt. The T2I transformation explicitly eliminates vocabulary limitations by converting a textual user query into a visual representation. This allows the search engine to retrieve relevant videos by computing visual similarities between the generated images and videos (within the same modality)—without needing to process query words explicitly.
The T2I transformation also addresses the complex query problem by representing search intentions spatially in a simplified image form, allowing the search engine to bypass reasoning on logical and spatial constraints, as these constraints are already visualized in the generated images. 
To further enhance robustness, we propose the I2T transformation, which paraphrases a user query by captioning the generated images using an image captioning model. This transformation ensures that the proposed framework remains effective in handling OOV and complex query problems, particularly when a search engine fails to interpret the generated images due to distortion and artifact issues—common challenges in image generation models \cite{rombach2021stablediffusion}. For instance, a distorted image of a person jumping, where the legs appear unnatural, could negatively impact retrieval performance. However, such image distortions produced by a diffusion model do not affect an image captioning model (e.g., BLIP-2 \cite{BLIP2}), which can generate a general description of the image. Moreover, generated captioning queries can help address the OOV problem because image captioning models share similar vocabulary knowledge with the search engine, as they are trained on similar captioning datasets. In addition to handling OOV issues, captioning queries can also decompose complex queries by describing the generated images from different viewpoints. 
For example, given the query \textit{Find shots of a red or blue scarf around someone's neck}, the I2T transformation generates sub-queries such as \textit{Find shots of a man wearing a red scarf}, \textit{Find shots of a girl wearing a red and blue scarf}, and \textit{Find shots of a person wearing a knitted red scarf}. Experimental results demonstrate that fusing these sub-queries for retrieval leads to higher retrieval performance.

Before sending the I2I, T2I, and I2T paraphrased queries to the search engine, we propose a consistency-based verification mechanism to eliminate factually incorrect queries caused by LLM hallucinations. This mechanism performs a novel verification by cross-checking the consistency between each paraphrased query and a set of QA pairs generated from the user query. Only those paraphrased queries that achieve the highest number of consistent QA pairs are retained.
The verified paraphrased queries are then submitted to the search engine for text-to-video and image-to-video retrieval. The results from these retrievals are fused to generate the final ranked list.
We have conducted extensive experiments for ad-hoc video search (AVS) and textual known-item search (KIS) on the TRECVid datasets \cite{Trecvid2016,V3C1}. The experiment results show that our proposed framework outperforms the state-of-the-art methods in terms of the mean xinfAP score for the ad-hoc video search task (i.e., Section \ref{AvS_comparison}) and the mean median rank for the textual known-items search task (i.e., Section \ref{tkis_exp}). Ablation studies are performed on the proposed consistency-based verification and query paraphrasing modules to evaluate their contribution to the whole framework (i.e., Sections \ref{exp_selfchecking_abstudy} and \ref{exp_query_transformers}).
We also provide empirical insights into how each transformation addresses the robustness, OOV, and complex query problems and how they complement the user query in term of retrieval performance and diversity (i.e., Section \ref{exp_query_transformers}). 

\section{Related Work}

\subsection{Text-to-Video Search}
Text-to-video originated from the evaluation in TRECVid \cite{trecvid2006}. The mainstream approaches have evolved from concept-based search \cite{Waseda2016, Jiang2007VIREO374L}, embedding-based search \cite{w2vvpp,vse, dualconding,LAFF}, to the more most interpretable embedding-based search \cite{dual_task,wu2023ITV,RIVRL} since the availability of many video caption datasets for cross-modal embedding feature and concept learning. However, the retrieval performance of the existing methods is hindered by the small size of the concept bank and the small amount of training data, and they are not always effective in directly answering a user query due to the out-of-vocabulary (OOV) problem. Some works have alleviated the OOV problem by building a huge concept bank \cite{wasedaAVS2020,WasedaMeiseiSoftbank2019}, exploring query expansion with WordNet hypernyms \cite{Waseda2018} for a better concept selection, and, most recently, generating over 7 million pseudo captions for web videos \cite{improvedITV}. These works inherently have the OOV due to the size and variety of the concept bank. Additionally, some studies \cite{Informedia2018, liMultiModalInductiveFramework2024a} have explored the use of external resources or knowledge to address the OOV problem. For instance, webly-labeled learning \cite{Informedia2018} learns OOV concepts by collecting related videos from the Internet. However, its effectiveness heavily depends on the retrieval accuracy of the Internet search engine. MMI-TVR \cite{liMultiModalInductiveFramework2024a}, on the other hand, leverages the broader knowledge of a multi-modal large language model (MLLM), such as LLaVA, to understand OOV concepts and let the LLM to determine whether they match salient visual-temporal concepts distilled from videos.

Besides the OOV problem, complex queries mixed with logical and spatial constraints are challenging to the existing methods. Only a few efforts try to address them, and the existing methods mainly focus on addressing one type of complex query. For example, the dual-task model \cite{dual_task} requires users to manually decompose a complex query with logical constraints into sub-queries, followed by combining the rank list of each sub-query by Boolean operations. BNL \cite{Understand_Negation_rumcc_MM} proposed to understand the logical operator \lq\lq NOT\rq\rq\ in video retrieval through negation learning by composing negative text-video pairs, which was implemented by inserting the word \lq\lq NOT\rq\rq\ before or after the verbs of the texts in positive text-video pairs. HGR \cite{hgr} addressed the spatial constraint in a user query by modelling the query and video through the events-actions-entities graph neural network. The improved ITV addressed the spatial constraint by adding compositional words to the concept bank and training the concepts using seven million pseudo-caption-video pairs \cite{improvedITV}.  

Different from the existing methods, in this paper, we propose to address the OOV and complex query problems through three verified transformations. 

\subsection{Query Paraphrasing and Expansion}
In video retrieval, query paraphrasing is rewriting a user query, and it works well with query expansion, which tries multiple queries to improve retrieval performance. Existing methods to have a new query are to replace the words in a user query with their synonyms \cite{queryParaphsingDocument} or create a new query based on the related candidates retrieved by the user query \cite{queryExpansionObjectRetrieval}. In the interactive video search, e.g., video Browser showdown (VBS), query paraphrasing is also widely used to retrieve targeting videos in a trial-and-error manner \cite{VBS2018}.

LLM-based T2T query paraphrasing has been a popular strategy in text retrieval \cite{wang2023querydoc} since the inception of chatGPT \cite{OpenAI2023ChatGPT} and other open-source LLM models, e.g.,  LLaMa  \cite{touvron2023llama}. For example, Query2doc expanded a user query with pseudo-documents generated by an LLM to improve text retrieval \cite{wang2023querydoc}. 
There are some attempts to use LLMs to paraphrase the given text in cross-modal representation learning \cite{LaCLIP}. LaCLIP rewrote a query using LLaMa  \cite{touvron2023llama} for data argumentation in training cross-modal representation and boosted the performances of the CLIP features \cite{CLIP} on multiple text-vision tasks. A few works have studied LLM-based query paraphrasing in video retrieval \cite{Waseda2023, VIREO_VBS2024}. ChatGPT was applied to make a small change on an AVS query to get 100 new queries to retrieve videos with diversity \cite{Waseda2023} and to rewrite the textual query in the textual KIS \cite{VIREO_VBS2024}. However, they have not studied the relationship between the paraphrased queries and the user query. They also have no mechanism to detect and remove those factually incorrect paraphrased queries. Nonetheless, our experiment (see Section \ref{exp_selfchecking_abstudy}) shows that having such verification can boost retrieval performance.

Paraphrasing a given text query by a text-to-image (T2I) transformation for video retrieval has seldom been explored as the quality of those T2I paraphrased queries cannot be guaranteed before the diffusion models and large pre-trained models become popular. Yanagi et al. \cite{genSceneImageForVideoSum} used T2I paraphrased queries for query-driven video summarization. Their work used a generative adversarial network (GAN) to perform T2I transformations for a given query sentence and measured the similarity of the T2I queries and video frames. They then concatenate those video frames most similar to the T2I queries as a video summary.  Yanagi et al. \cite{genSceneImageForVideoSum} transformed a textual query into a visual query for image retrieval. However, their visual query was image regions extracted from the images. Furthermore, in their study, the paraphrased visual query could not bring significant gains for retrieval in their experiment. Until recently, Ma et al. explored the T2I transformations in the KIS task \cite{VIREO_VBS2024}. However, the performance improvements are not consistent on query sets in different years due to generation without verification.

To our knowledge, there is no study on paraphrasing a user query by an image-to-text (I2T) transformation for video retrieval. However, our empirical study has proved that the I2T query paraphrasing is essential in safeguarding the proposed framework for solving the OOV and complex query problems. Moreover, its retrieval performance has outperformed the T2I transformation even though it is without self-verification in our experiments. With self-verification, it outperforms T2I by a larger margin.

\subsection{QA in Video Search}
We also review closely related prior works \cite{askAndConfirm_ICCV,QAforRetrieval_MM,liang2023simplebaselineQA} on using the concept of Question-and-Answer (QA) in video retrieval. These works mainly focused on using QA to remove the ambiguity in user queries for interactive video retrievals and improve the retrieved rank lists by interacting with users.  

There are major differences between ours and the prior works. First, the prior works (except \cite{liang2023simplebaselineQA}) were semi-automated, which relied on humans to obtain the answer to each question, whereas our design factors out this human aspect, thereby being more scalable due to full automation. 
Second, their QA mechanism has no quality control over the answers. In contrast, ours is to generate QA pairs, check the consistency among them for each paraphrased query, and eliminate those queries that are less consistent.

\section{LLM-based Query Paraphrasing for Video Search}

\begin{figure*}[t]  
\centering  
  \centering  
  \includegraphics[width=0.9\linewidth]{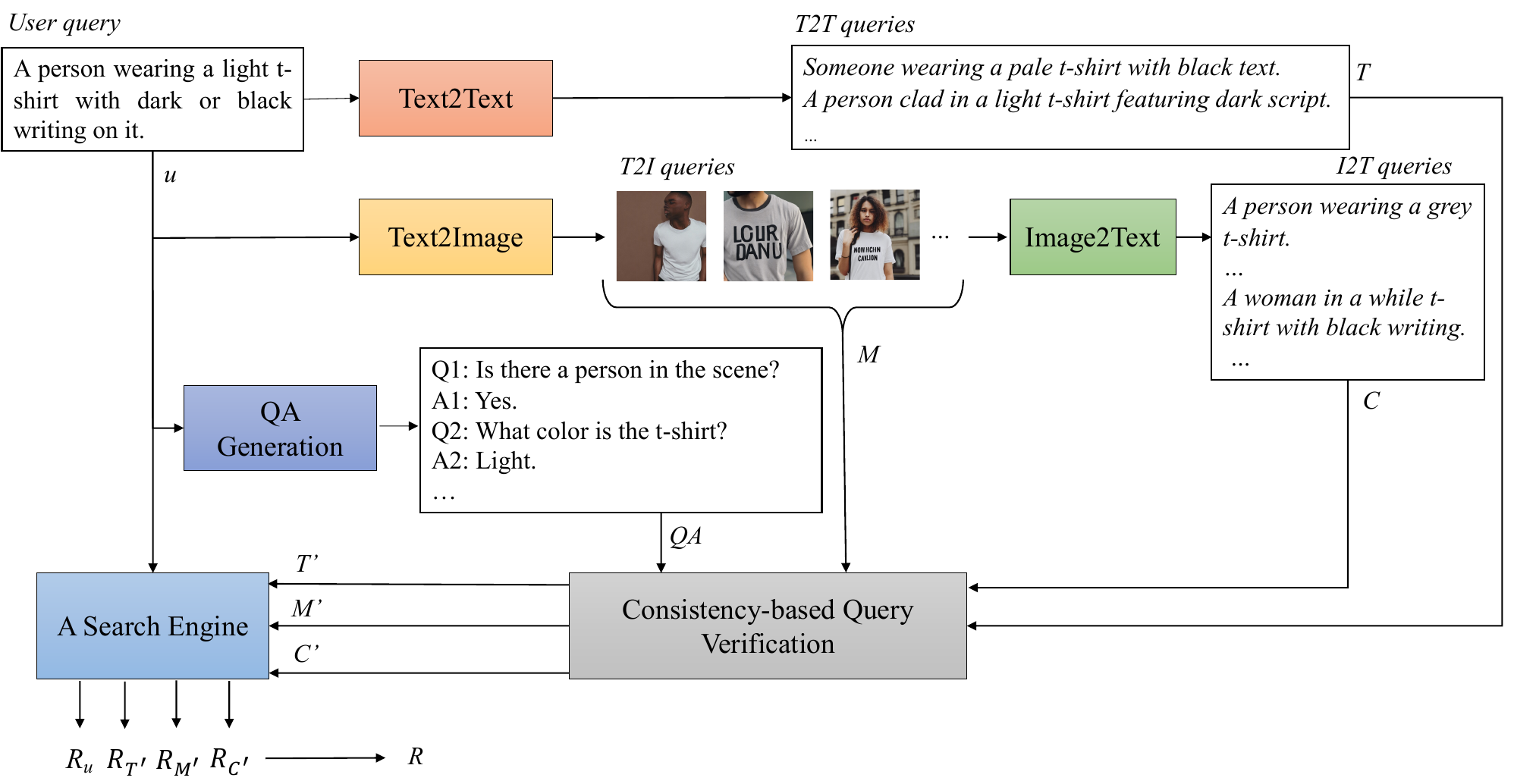}  
    \caption{The proposed LLM-based query paraphrasing framework for video search. The user query is rephrased by Text2Text, Text2Image, and Image2Text transformations, followed by consistency-based verification and then video retrieval through a search engine. The final rank list is obtained by the ensemble of the four intermediate rank lists.}   
    \label{fig:GENsearch_framework}
\end{figure*}

In this section, we present our proposed LLM-based query paraphrasing framework. Fig. \ref{fig:GENsearch_framework} shows the overview of the proposed framework. It consists of three steps to retrieve videos. The first step is query paraphrasing through three transformations, which transfers the user query to I2I, I2T, and I2T queries by text-to-text, text-to-image, and image-to-text transformations, respectively. The second step is paraphrased query self-verification, which verifies the rephrased queries by checking the consistency against QA pairs generated from the user query to validate their faithfulness to the user query. The final step is query-to-video search, including user-to-video search, T2T-to-video search, T2I-to-video search, and I2T-to-video search. The user query and the verified paraphrased queries are sent to the search engine (i.e., the improved ITV model) to obtain four rank lists of videos. These rank lists are then fused into the final video retrieval result.

\subsection{LLM-based Query Paraphrasing}

Capitalizing on the pre-trained generative models (e.g., LLM \cite{OpenAI2023ChatGPT} and stable diffusion \cite{rombach2021stablediffusion}), we propose to understand user queries through multiple paraphrasing dimensions: text-to-text (T2T), text-to-image (T2I), and image-to-text (I2T) transformations.

The T2T transformation passes a textual user query $u$ to a pre-define prompt, i.e., $T2Tprompt(x,n)$=\lq\lq rewrite a user query '$x$' $n$ times\rq\rq, to a LLM model (i.e., chatGPT4 \cite{openai2023gpt4}) to produce $n_T$ number of T2T paraphrased queries as:
\begin{equation}
\label{eq:T2T_tramsformation}
T = \{t_1,...,t_{n_T}\}= LLM(T2Tprompt(u,n_T)). 
\end{equation}

The T2I transformation sends a textual user query to a pre-trained text-to-image generation model (i.e., stable diffusion (SD--T2I) \cite{rombach2021stablediffusion}) and generates $n_M$ number of visual queries as:
\begin{equation}
\label{eq:T2T_tramsformation}
M = \{m_1,...,m_{n_M}\}= SD-T2I(u).
\end{equation}

Besides, we also propose to include I2T transformations to safeguard the ability of the proposed framework to deal with OOV and complex query problems in case the T2I queries have serious image distortion problems resulting in bad alignments with videos. The I2T transformation takes a T2I query $m_i$ as an input to an image captioning model (i.e., BLIP-2 (BLIP2-IC) \cite{BLIP2}) to generate $n_C$ number of I2T queries as:
\begin{equation}
\label{eq:T2T_tramsformation}
C = \{c_1,...,c_{n_C},...\}= \{BLIP2-IC(m_i)\}^{n_M}.
\end{equation}

\subsection{Verifying Paraphrased Queries as Valid}
\label{sec_QA_verification}

\subsubsection{Motivation and Idea}
T2T, T2I, and I2T transformations on a user query are imperfect.
For instance, an image query may only be partially consistent with the user query due to the hallucination of a generative model. A paraphrased query through captioning an image query may depict the image query from a viewpoint different from the user query's. A textual query rewritten from the user query could miss some subject-of-interest of the user query. 

We propose to verify the faithfulness of the transformed queries before sending them to a search engine for video retrievals by checking each such query with the QA pairs produced from the user query.
For instance, the complex user query \textit{Find shots of people dancing or singing while wearing costumes outdoors} specifies requirements on the subject, action, object, and location, and a T2I transformation may imperfectly produce an image query for it.
If simply asking an LLM (e.g.,  LLaMa 3 \cite{touvron2023llama}) whether such an image query faithfully represents the user query (which, for instance, may overlook the requirements such as the ``outdoors'' locational information), unfaithful image queries such as those showing people dancing indoors may pass such a question-and-answer strategy.

We propose to use an aspect-oriented approach to address this issue. We will generate QA pairs from a user query to verify each aspect of such an image query: person/being, action, object, location, time, colour, and quantity. Only those image queries with the maximal number of matched QA pairs are regarded as passing the verification in our framework, are considered to be the most trustworthy, and are labelled as \textit{valid}.

\subsubsection{Generation of Question-Answer Pairs}
We propose to generate question-answer (QA) pairs for a given user query using an LLM (i.e., chatGPT4 \cite{openai2023gpt4}). Specifically, there are two kinds of QA: open-vocabulary QA and YES/NO QA, and there are seven aspects: person, object, action, location, time, colour, and quantity. We design the prompt for the LLM to be 
$GenQAprompt(x)$=\lq\lq You act as a question generator. Generate various simple and short QA pairs (including open-vocabulary QA and YES/NO QA) for a given query, with extra emphasis on person/being, action, object, location, time, colour, and quantity when specified. The given query is '$x$'\rq\rq\ to generate multiple QA pairs for each user query $u$ as:
\begin{equation}
\label{eq:T2T_tramsformation}
QA = \{(q_1,a_1),...,(q_{n_u}, a_{n_u})\}= LLM(GenQAprompt(u)). 
\end{equation}

\subsubsection{Validating Paraphrased Queries}
After obtaining the QA pairs, we verify each paraphrased query against them by asking
an LLM (which is chatGPT4 \cite{openai2023gpt4} in our experiment) to return the number of QA pairs that are correct in the context of the query. 

We design the prompt for the query verification as ValidQprompt $(x_i,z)$="Given an image (or a text) and a set of QA pairs, verify if the image (or the text) is aligned with each QA pair. Output the verification result for each pair and count the number of aligned QA pairs. The given image (or text) is '$x_i$', and the QA pairs are listed as follows '$z$' ", in which $x_i$ is a T2T query $t_i$, a T2I query $m_i$, or an I2T query $c_i$.
\begin{equation}
\label{eq:T2T_tramsformation}
{y_i} = LLM(ValidQprompt(x_i, QA)),
\end{equation}
where $y_i$ is the number of aligned QA pairs for a paraphrase query $x_i$. In each transformation, those paraphrased queries that are associated with maximal numbers of LLM-returned QA pairs are treated as valid, producing three valid paraphrased query sets $T'$, $M'$, and $C'$, which contain the valid T2T, T2I, and I2T queries, respectively.

\subsection{Retrieving Videos Using Valid Queries}

The valid queries are subsequently sent to a search engine for video retrieval. In our experiments, we use a recent text-to-video model (i.e., improved ITV \cite{improvedITV}) as our search engine. The improved ITV model consists of a visual encoder and a textual encoder to encode a video and a text query in a joint latent space, respectively. It also has a visual-textual concept decoder to interpret the latent features into concept vectors. It has proposed a fusion search, which is a linear combination of concept-based search and embedding-based search. 

In this paper, we perform text-to-video retrieval by using the fusion search of the improved ITV. Furthermore, we perform image-to-video retrieval based on the visual encoder of improved ITV. Specifically, a T2I query is processed as a one-frame video and sent to the visual encoder to extract a visual embedding feature. A visual similarity rank list is returned based on the similarities of the visual embedding feature of the T2I query and the visual embedding features of test videos. We follow other methods \cite{dual_task,improvedITV} to use cosine similarity to measure the distance of two embedding features. In general, we get a score list for all the test videos from a search engine by inputting a query $x$ as:
\begin{equation}
\label{eq:search}
S_x= SearchEngine(x)
\end{equation}
For each valid paraphrased query, we get a score list from a search engine. For each transformation, we perform an average ensemble of all the valid queries with equal weight. As a result, we have one score list from each transformation, i.e., $S_{T'}= norm(\frac{1}{|T'|}\sum_{i=1}^{|T'|}S_{t'_i})$ from the T2T, $S_{M'}= norm(\frac{1}{|M'|}\sum_{i=1}^{|M'|}S_{m'_i})$ from the T2I, $S_{C'}= norm(\frac{1}{|C'|}\sum_{i=1}^{|C'|}S_{c'_i})$ from the I2T transformations, respectively. We also perform a linear ensemble on the score lists of the user query (i.e.,  $S_{u}$) and of three transformations (i.e., $S_{T'}$, $S_{M'}$, and $S_{C'}$) to get the final result as: 
\begin{equation}
\label{eq:ensemble}
\begin{split}
S=&ensemble(S_u, S_{T'}, S_{M'}, S_{C'}) \\=&S_u+S_{T'}+0.5*S_{M'}+ S_{C'}
\end{split}
\end{equation}
\begin{equation}
\label{eq:ensemble}
R=argsort(S)
\end{equation}
We use the weights of 1 for $S_u, S_{T'}$, and $ S_{C'}$ and 0.5 for $ S_{M'}$ in the linear ensemble. We give a lower weight to T2I queries by considering images' inherent limitations in modelling multiple motions. Finally, we rank all the test videos based on $S$ and get the final rank list $R$. In the following sections, we name our framework as GenSearch.

\section{Experiments}

\subsection{Experimental Settings}




\label{exp_setting}
\textbf{Datasets}. 
The evaluations are conducted on three TRECVid AVS datasets: IACC.3 \cite{Trecvid2016}, V3C1 \cite{V3C1}, and V3C2 \cite{V3C1}. IACC.3 has 335k test videos, V3C1 has 1 million videos, and V3C2 has 1.4 million videos. 
Each year, TRECVid releases a query set for the AVS task, which consists of 20 or 30 queries. IACC.3, V3C1 and V3C2 were used in 2016-2028, 2019-2021 and 2022-2023, respectively.
Similarly, each year, VBS releases a query set for the textual KIS (TKIS) task, which has 6-12 queries. Specifically, V3C1 was used in 2019-2021, and the combination of V3C1 and V3C2 was used in 2022-2024 for the KIS task.   
We evaluate GenSearch on eight years of AVS query sets (from tv16-tv23) and six years of KIS query sets (from tkis-19 to tkis-24).

\textbf{Evaluation metric}. We use extended inferred average precision (xinfAP) \cite{Trecvid2019} with a search length of 1000 to evaluate AVS results and median rank (medR) for TKIS. To assess the overall performance of the search system, we report the mean xinfAP/medR across all test queries. Additionally, we use the mean pairwise distance of a sub-sampling set to measure the diversity of the retrieval results.

\textbf{Implementation details}. 
For the T2T transformation, we use the  chatGPT4 to rewrite a user query 10 times. For the T2I transformation, we use the stable diffusion  \cite{rombach2021stablediffusion} with different initial variables to generate images, i.e., seeds = $[$10, 100, 1000, 10000, 100000$]$. Three images are produced for each seed, thereby producing 15 images for each query. We adopt the default values for other parameters. For the I2T transformation, we use the BLIP-2 image captioning model \cite{BLIP2} to generate ten captions for each image query. We use the improved ITV  \cite{improvedITV} as the search engine. The implementation settings of improved ITV are the same as \cite{improvedITV}.
Note that our proposed framework can be a plug-in module for any video search engine, thus we also report the results based on BLIP-2 model to test the robustness of our method. 

\subsection{AVS Performance Comparison}

\label{AvS_comparison}

\begin{table}[t]
\caption{Performance comparison on TRECVid AVS datasets. The reproduced results are marked with *, while the unavailable results are indicated by /. } 
\label{tab:tenSearch_avs_comparison}
\centering
\resizebox{\linewidth}{!}{
\begin{tabular}{l|ccc|ccc|cc|c}
    \toprule
Datasets                     & \multicolumn{3}{c|}{IACC.3}               & \multicolumn{3}{c|}{V3C1}   & \multicolumn{2}{c|}{V3C2}   \\
\hline
Query sets                         & tv16 (30)         & tv17 (30)           & tv18  (30)          & tv19 (30)   &tv20 (20) &tv21 (20) &tv22 (30) &tv23 (20)  &Mean     \\
                          \hline
\multicolumn{5}{l}{TRECVid top results:}                                                  \\
Rank 1     & 0.054  & 0.206  & 0.121 & 0.163   &0.359   &0.355   &\textbf{0.282} &0.292\\
Rank 2     & 0.051   & 0.159   & 0.087  & 0.160    &0.269   & 0.349   &0.262 &0.285   \\
Rank 3     & 0.040   & 0.120   & 0.082       & 0.123     &0.229     &0.343 &0.210 &0.272 \\
\hline
\multicolumn{9}{l}{Concept-based Search:}  \\
ConBank (auto)  & /  & 0.159 \cite{Waseda_Meisei2017}  & 0.060 \cite{Waseda2018}        & /    & /     & /     & /  & /  & /\\
ConBank (manual)   & 0.177 \cite{Waseda2016}   & 0.216 \cite{Waseda_Meisei2017}  & 0.106 \cite{Waseda2018}        & 0.114  \cite{WasedaMeiseiSoftbank2019}  & 0.183 \cite{wasedaAVS2020}    & /  & /  & /    & /\\
 \hline
\multicolumn{9}{l}{Embedding-based Search:}  \\
VideoStory ~\cite{videostory} & 0.087          & 0.150           & /              & /          & /   & /    & /  & /    & /     \\
VSE++ ~\cite{vse}           & 0.135          & 0.163          & 0.106         & /    & /     & /       & /    & /    & /  \\
W2VV ~\cite{w2vv}                      & 0.149          & 0.198          & 0.103          & /  & /   & /   & /    & /    & /   \\  
W2VV++ ~\cite{w2vvpp}       & 0.150         & 0.207           & 0.099         & 0.146     & 0.199      & /    & /    & /    & /   \\
W2VV++ variant ~\cite{W2VVpp_Case_Study}       & 0.156          & 0.218           & 0.110          & 0.151     & /    & /    & /    & /    & /  \\
Dual coding ~\cite{dualconding}              & 0.160          & 0.232           & 0.120         & 0.163  & 0.208    & /     & /    & /    & /  \\ 
Dual coding attention ~\cite{dual_coding_improved}              & 0.159          & 0.244           & 0.126         & /    & /     & /    & /    & /    & / \\ 

HGR ~\cite{hgr}             &/         & /           & /       & 0.142   &0.301     & /    & /           & /    & /     \\ 
SEA ~\cite{tmm2021-sea}             & 0.164         & 0.228           & 0.125       & 0.167     & 0.186  & /    & /           & /  & /  \\ 
CLIP  \cite{CLIP}       &   0.182 & 0.217    & 0.089  &0.117   & 0.128  &  0.178  &0.124 & 0.109 &0.143 \\
CLIP4CLIP \cite{Luo2021CLIP4Clip} & 0.182 &0.217 &0.089 &0.133 & 0.149 &0.188 & 0.121 & 0.109 &0.149\\
BLIP-2 \cite{BLIP2}     &0.213 &0.226 & 0.168&0.199 &0.222 &0.273 &0.164 & 0.203 &0.209 \\

LAFF* \cite{LAFF}             &    0.188   &   0.261        &  0.152   &  0.215     &  0.299   &  0.300  & 0.178 &0.172  &0.221\\ 

ChatGPT-QE \cite{Waseda2023} & /&/&/&/&/&/ &0.112&/ & /\\
LanguageBind  \cite{zhu2024languagebind} & 0.182&0.238&0.145& 0.162 &0.221 &0.254 &0.184&0.146 &0.192\\
GLSCL \cite{zhangTextVideoRetrievalGLSCL2025}  &0.165 &0.189 & 0.086 &0.142 & 0.206 & 0.182& 0.127&0.119 &0.152\\
 \hline
\multicolumn{9}{l}{Fusion of Embedding-based and Concept-based Search:}  \\

Dual-task ~\cite{dual_task}              & 0.185          & 0.241           & 0.123        &0.185    & /   & /    & /           & / & /\\ 
Hybrid space ~\cite{tpami21-dual-encoding}              & 0.157          & 0.236           & 0.128         & 0.170    & 0.191  & 0.162    & /           & /  & /\\
RIVRL*~\cite{RIVRL}  & 0.159          & 0.231           & 0.131        & 0.197    & 0.278  & 0.254 & 0.179  & 0.177 &0.201 \\
ITV \cite{wu2023ITV}                &  0.211    & 0.292   &   0.170  &   0.227          &0.345  & 0.318 & 0.150  & 0.170 &0.235 \\
Improved ITV \cite{improvedITV} & 0.280 &	0.349 &	0.165 &	0.242	&0.352	&0.365 &	0.235&	0.295 &0.285\\
\hline
\multicolumn{5}{l}{The proposed models:}  
                               \\
                   
GenSearch (on BLIP-2 \cite{BLIP2})          & 0.237	& 0.258	&\textbf{0.180}	&0.226	&0.259	&0.292	&0.212	&0.233 &0.237\\
 
GenSearch (on Improved ITV \cite{improvedITV})          & \textbf{0.281}	&\textbf{0.366}&\textbf{0.180}	&\textbf{0.263}	&\textbf{0.359}	&\textbf{0.377}	&0.258	&\textbf{0.311} &\textbf{0.299}  \\

\bottomrule
\end{tabular}
}
\end{table}

We compares GenSearch with the existing methods and the best results on TRECVid for the corresponding query sets. Specifically, several kinds of AVS methods are inlcuded for comparison, i.e.,
concept-based, embedding-based methods and the fusion of concept-based and concept-free searches. 
For the recent approaches, LAFF \cite{LAFF} RIVRL \cite{RIVRL}, improved ITV \cite{improvedITV} and GLSCL \cite{zhangTextVideoRetrievalGLSCL2025}, we reproduce their results using the same setting as ours for fair comparisons (e.g., the same training sets and visual features) based on their published codes.


\begin{figure*}[]  
\centering  
  \centering  
  \includegraphics[width=1\linewidth]{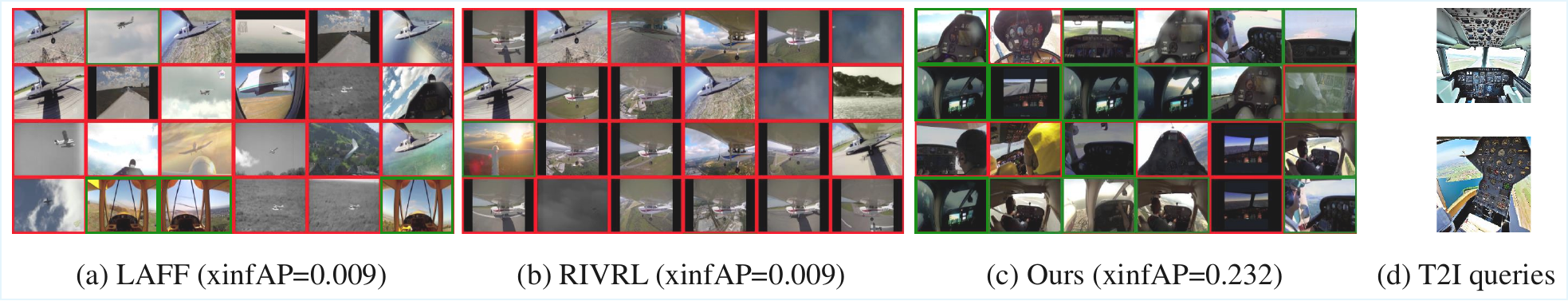}  
    \caption{Retrieved videos on the query-639 \textit{Find shots for inside views of a small aeroplane flying} by LAFF in (a), RIVRL in (b), and our in (c). Videos with red and green border indicates correct and wrong results, respectively.}   
    \label{fig:insidePlane}
\end{figure*}

\begin{figure*}[]  
\centering  
  \centering  
  \includegraphics[width=1.01\linewidth]{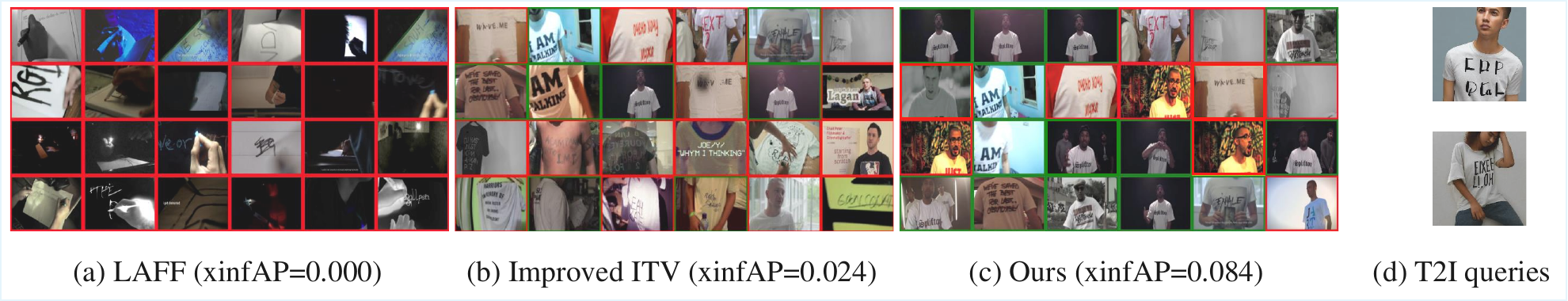}  
    \caption{Retrieved videos on the query-710 \textit{Find shots of a person wearing a light t-shirt with dark or black writing on it} by LAFF in (a) and the Improved ITV in (b), and our in (c).}   
    \label{fig:lightshirt}
\end{figure*} 

Table \ref{tab:tenSearch_avs_comparison} lists the comparison results. GenSearch (built upon the improved ITV model) outperforms all existing methods by a large margin on all query sets except tv16 and tv20. Specifically, it shows better performances on 165, 181, and 128 out of 210 test queries in the comparisons with three state-of-the-art methods, LAFF \cite{LAFF}, RIVRL \cite{RIVRL}, and the improved ITV \cite{improvedITV}, respectively. Consistent improvement is also observed when comparing GenSearch (built upon the BLIP-2 model) with BLIP-2 with a mean xinfAP of 14.23\% across query sets. We found that GenSearch improves state-of-the-art methods on their bad-performing queries (i.e., those with xinfAP $\leq 0.1$). Specifically, LAFF \cite{LAFF}, RIVRL \cite{RIVRL}, and the improved ITV \cite{improvedITV} have 78, 103, and 58 bad-performing queries, respectively, and GenSearch has improvements on 80.77\%, 86.4\%, 51.72\% of them, respectively. Moreover, GenSearch boosts the improved ITV, which has the same training set and concept bank, on OOV queries by 14.21\%. For example, for the query-679 \textit{Find shots of a ladder with less than 6 steps}, where the concept \textit{less than 6 steps} is not in the vocabulary, GenSearch outperforms the improved ITV by 48.84\% through text-to-visual transformation.
GenSearch also has a higher performance than the TRECVid top-1 results on all query sets except tv22. The TRECVid top results usually use various models in the ensemble to boost the retrieval performance. For example, the top-1 solution on the tv22 query set combined the results of more than 100 rank lists from five models. Randomization test \cite{randomization_test} shows that GenSearch built on top of the improved ITV significantly outperforms all the compared methods with $p$-value $\leq$ 0.01.

\begin{figure}[]  
\centering  
  \centering  
  \includegraphics[width=1\linewidth]{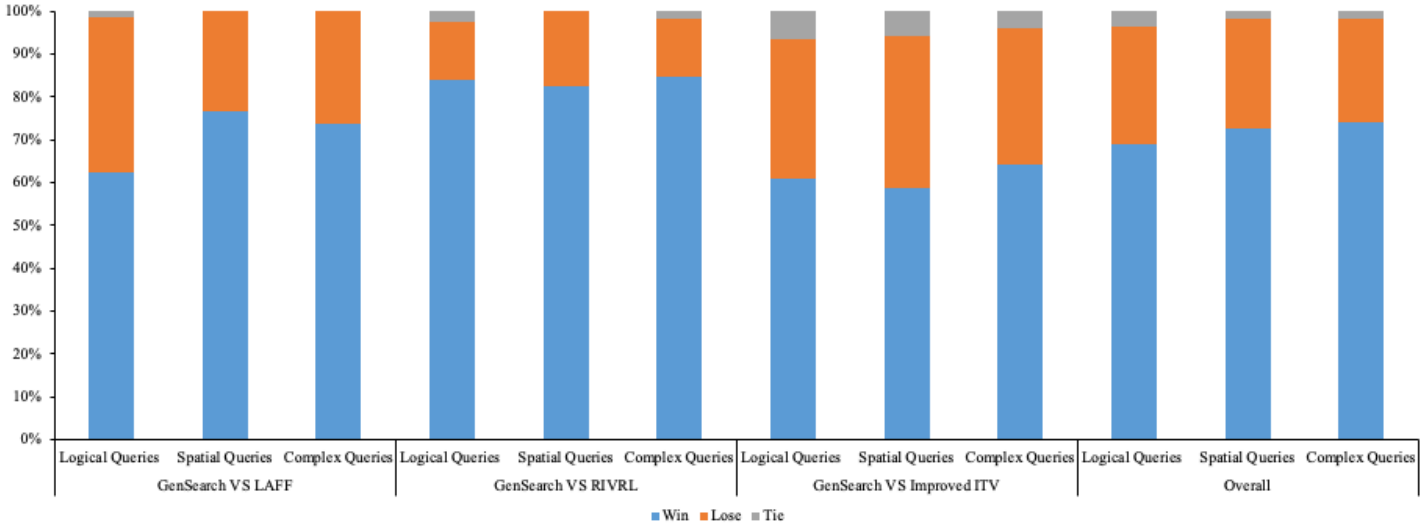} 
    \caption{AVS competitions with LAFF, RIVRL, and improved ITV methods on the logical, spatial, and complex queries. The percentage of queries won (in blue), lost (in orange), and tied (in gray) by GenSearch is shown for each competition.}   
    \label{fig:SpecifySquery}
\end{figure} 

We further compare GenSearch with the state-of-the-art methods on specified queries, i.e., queries containing at least one logical operator (i.e., AND, OR, NOT) (logical queries), queries that have spatial constraint (e.g., \textit{in front of} and \textit{behind}), and complex queries defined by TRECVid. Specifically, TRECVid has classified 210 queries into 13 different complexity based on the composition of four elements: person, action, object, and location. We treat a query containing more than two elements as a complex query, and there are 104 complex queries in total. Fig. \ref{fig:SpecifySquery} shows how many queries the GenSearch win, lose and tie in the competitions. In general, GenSearch has an average of 71.86\% to win across all compared methods over all queries. Specifically, for 74 logical queries, GenSearch has higher retrieval rates on 62.16\%, 83.78\%, and 60.81\% of queries than LAFF, RIVRL, and the improved ITV, respectively. For example, for the query-627 \textit{Find shots of a person holding a tool and cutting something}, the improvement gains towards LAFF, RIVRL, and the improved ITV are 66\%, 172\%, and 160\%, respectively. In term of spatial query, GenSearch has at least 58.82\% to beat the compared methods. For example, for the query-729  \textit{Find shots of A ring shown on the left hand of a person}, LAFF, RIVRL, improved ITV \cite{improvedITV} struggle on the spatial constraint of \textit{left hand}, while GenSearch resolves this constraint by T2I transformation. Besides, GenSearch shows great advantages on solving complex queries. Among 104 queries, GenSearch outperforms LAFF \cite{LAFF}, RIVRL \cite{RIVRL}, and improved ITV \cite{improvedITV} on 76, 86, and 66 queries, respectively. 

Fig. \ref{fig:insidePlane} and Fig. \ref{fig:lightshirt} further visualizes how GenSearch outperforms existing methods (LAFF, RIVRL, and Improved ITV) on complex video retrieval queries involving spatial constraints and logical reasoning. For query-639 \textit{Find shots for inside views of a small aeroplane flying}, LAFF and RIVRL focus on the aeroplane flying while neglecting the spatial constraint of the cabin view. In contrast, GenSearch visualizes the input textual query to images by T2I transformation and further transforms the images to various captioning queries with more related context, e.g., \lq\lq A cockpit view of a small airplane engine with various controls and gauges\rq\rq, and \lq\lq Two people are sitting in the cockpit of a small airplane \rq\rq. For query-710 \textit{Find shots of a person wearing a light t-shirt with dark or black writing}, it encounters OOV problem because the concept \textit{light t-shirt} rarely appears in the training data and the concept bank. Besides, this query contains the logical word \textbf{OR}. LAFF cannot understand the query well and find video shots of writings on paper at dark. With over 7 million text-video pairs for pre-training, the improved ITV can retrieve some true positives but it fails at understanding all the requirements of query-710. Instead, GenSearch finds more relevant video shots by having related visual queries as shown in Fig. \ref{fig:lightshirt}(c). Moreover, the I2T queries, e.g., \lq\lq A man in a white t-shirt with black writing\rq\rq\ and \lq\lq A woman in a light-coloured shirt with a dark message on it\rq\rq, decompose the complex query into simple sub-queries, which relax the requirement of logical reasoning of the search engine.

\subsection{Textual KIS Performance Comparison}
\label{tkis_exp}

\begin{table}[t]
\caption{Textual-KIS performance comparison (MedR).
}
\label{table:textual-KIS_comparison}
\centering
\resizebox{\linewidth}{!}{
\begin{tabular}{l|ccc|ccc|c}
    \toprule
                       Datasets                      & \multicolumn{3}{c|}{V3C1} & \multicolumn{3}{c|}{V3C1+V3C2}  &  \\
                       \hline
                       Query sets  & tkis-19 (12)  & tkis-20 (12)  & tkis-21 (10)  & tkis-22 (12) & tkis-23 (12) & tkis-24 (6)  & Average  \\
                       \hline
                       BLIP-2 \cite{BLIP2} & 
                       \textbf{1.5 }&
4.5&
52&
8&
114&
\textbf{9.5}&
31.58
\\ 
                       LAFF \cite{LAFF} &
                       5.5&
26&
4.5&
66&
513&
15&
105
\\
                       RIVRL \cite{RIVRL}  &
                       9&
29&
116&
191.5&
594.5&
11&
158.5

\\
 Improved ITV  \cite{improvedITV}
&8 &
\textbf{2.5}&
4&
24.5&
19&
14.5&
12.08
\\ 


Manually query \cite{VIREO_VBS2024}  
&/
&/
&/
&60.5
&43
&/
&51.75

\\

GLSCL \cite{zhangTextVideoRetrievalGLSCL2025} & 9 & 28 & 48 & 60 & 1,985 &85 & 369.17\\
                       \hline
                       GenSearch  & 
                       \underline{3.5}&
\underline{3}&
\textbf{2.5}&
\textbf{3.5}&
\textbf{15}&
\underline{10.5}&
\textbf{6.33}
 \\

  \bottomrule
\end{tabular}
}
\end{table}

We also verify the ability of the proposed GenSearch framework on the Textual-Known-item search (TKIS) task. Comparing with AVS task, TKIS task has longer query (i.e.,  around 48 words Verses 9 in AVS tsak). Besides, a TKIS query is only align with one ground truth video and the test video corpus is usually very large. For example, more than one million video shots are in V3C1 \cite{V3C1} and more than 2 million video clips are in the combination of V3C1 and V3C2). 

\begin{figure*}[]  
\centering  
  \centering  
  \includegraphics[width=1\linewidth]{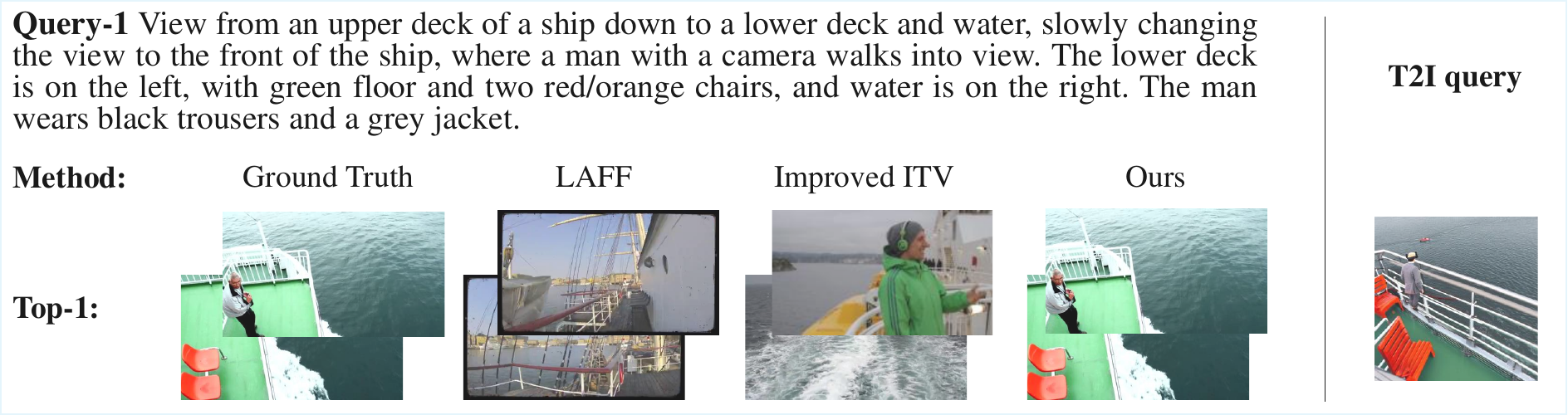}  
    \caption{Left: Top-1 retrieved video comparison with LAFF \cite{LAFF} and Improved ITV \cite{improvedITV} on a T-KIS query. Right: examples of T2I queries.}   
    \label{fig:TKIS-t22-11}
\end{figure*}

Table \ref{table:textual-KIS_comparison} shows that GenSearch obtains the lowest mean median rank across six-year query sets on two huge video datasets and outperforms both the widely used vision-text representation learning models and the state-of-the-art text-video retrieval model on most query sets. Moreover, it also outperforms the results obtained by the manually formulated queries based on a same search engine. Specifically, GenSearch obtains better retrieval ranks on 61\%, 44\%, 58\% of queries in comparison with CLIP, Improved ITV, and maually formulated queries, respectively. Meanwhile, there are 25\%, 33\%, 17\% tie queries, respectively, in the comparisons.   Fig. \ref{fig:TKIS-t22-11} visualizes the comparison with recent methods. GenSearch manages to improve the rank the ground truth to top 1 due to precision translation of OOV textual query to visual query, In contrast, LAFF and Improved ITV models fail to capture green deck and red chairs because of the OOV problem.

\subsection{Ablation Studies}
\label{ablation_study}

\begin{table}[]
\caption{AVS performance comparison before and after applying the QA verification on the paraphrased queries.}
\label{exp:verification_comparison}
\centering
\resizebox{\linewidth}{!}{
\begin{tabular}{l|ccc|ccc|cc|c}
    \toprule
                       Datasets                      & \multicolumn{3}{c|}{IACC.3} & \multicolumn{3}{c|}{V3C1} & \multicolumn{2}{c|}{V3C2} &  \\
                       \hline
                       Query sets               & tv16  & tv17  & tv18  & tv19 & tv20 & tv21  & tv22 & tv23 & Mean  \\
                       \hline
      T2T-to-video search: &       &       &       &       &  &&&&     \\
                        before  verification  &  0.276 &	\textbf{0.346}	&\textbf{0.160} &	0.238	&\textbf{0.348} &	0.364	&0.213	&0.296 &0.280\\
                          after  verification:     &\textbf{ 0.282}      &   0.343    & 0.159      &  \textbf{0.240}     & 0.346 & \textbf{0.374}& \textbf{0.221}& \textbf{0.297}&  \textbf{0.283}   \\
                           \hline

                           
                       T2I-to-video search: &       &       &       &       &  &&&&     \\
                    before  verification & 0.201&	0.232&	0.120&	0.214&	0.246&	0.260	&0.205	&0.228	&0.213\\
                    after  verification& \textbf{0.207}&	\textbf{0.236}	&\textbf{0.128}	&\textbf{0.215}	&\textbf{0.256}	&\textbf{0.270}	&\textbf{0.216} & \textbf{0.232}&	\textbf{0.220}\\
                       \hline
                       I2T-to-video search:     &       &       &       &       &  &&&&     \\
                     before  verification & 0.226&	0.288	&0.170	&0.209&	0.270	&0.279&	0.188&	0.256	&0.236\\
                    after  verification & \textbf{0.233} &	\textbf{0.328}&	\textbf{0.185}	&\textbf{0.214}&	\textbf{0.304}	&\textbf{0.323}&	\textbf{0.221}&	\textbf{0.282} &\textbf{0.261}\\

  \bottomrule
\end{tabular}
}
\end{table}

\subsubsection{Impact of Consistency-based Verification}
\label{exp_selfchecking_abstudy}
We first report the evaluation of the effectiveness of the proposed QA verification on the paraphrased queries. Specifically, we compare the AVS performances on the TRECVid query sets before and after the verification. In other words, the only variable here is whether the query verification model participates in the search process. The results are shown in Table \ref{exp:verification_comparison}. As most of the T2T queries are faithful to the original queries, the results before and after verification are very similar on most query sets. Specifically, the performances of 164 out of 210 AVS queries are unchanged. 35 and 11 queries are improved and dropped, respectively. For the T2I-to-video search and I2T-to-video search, the performance improvements are consistent on all query sets. Specifically, for the T2I-to-video search, the performances of 76, 90, and 44 AVS queries are unchanged, improved, and dropped after verification, respectively. The performance boost by verification is most significant in the I2T-to-video search because I2T queries inherently could have much more noise due to different viewpoints when describing an image. Randomization test \cite{randomization_test} shows that the verification process significantly improves the retrieval performances on T2I- and I2T-to-video search at $p$-value$<=0.01$.

\subsubsection{Retrieval Performance Comparisons of the Query Transformations}
\label{exp_query_transformers}

In this section, we evaluate the effectiveness of the user, T2T, T2I, and I2T queries in retrieving videos. The experiments are conducted based on Eq. (\ref{eq:ensemble}), and we compare the effectiveness and diversity of using them along or their combination in retrieving videos. 

Tables \ref{exp:comparison_original_query_generative_query} and \ref{exp:comparison_original_query_generative_query_on_specificType} compare their retrieval performance on general types of queries (i.e., all queries of three TRECVid AVS datasets) and on specific types of queries from the AVS datasets, respectively. In general types of queries, the user query beats the paraphrased queries in five out of eight query sets with the highest mean xinfAP 0.285. Specifically, the T2T query shows competitive performance with the user query on most query sets except tv22. The I2T query has significantly better performances than other queries in sets tv18 and tv22. However, when we compare the paraphrased queries and the user query on specific types, i.e., logical, OOV, and spatial-constrained queries, as shown in Table \ref{exp:comparison_original_query_generative_query_on_specificType}, I2T achieves the highest retrieval rate and beats the user query by 11.7\% in general. T2T and I2T queries significantly outperform the user query on negation, OOV and spatial types. For example, for query-730  \textit{A man is holding a knife in a non-kitchen location}, I2T paraphrases the out-of-vocabulary concept \lq\lq a non-kitchen location\rq\rq\ to \lq\lq woods\rq\rq, the performance is improved by 124.16\% and T2T improves the performance of query-569\textit{ Find shots of people standing in line outdoors} by replacing the out-of-vocabulary concept \lq\lq standing in line\rq\rq\ to \lq\lq lineup\rq\rq\ which is a concept in the concept bank. In contrast, the T2I query cannot compete with the other three, either in terms of general type or specific type. We found that the bad performance of the T2I query could be attributed to the existing way of performing T2I-to-video search, which mainly relies on semantic similarity on the whole image without query-specific attention. Therefore, for those queries that require fine-grained details, the T2I query fails miserably. For example, for the query-745 \textit{ Find shots of a person wearing gloves while biking}, T2I queries visualize the user query correctly, but the retrieved videos are all about a person riding a bike with or without gloves.

\begin{table}[]
\caption{AVS performance comparison between the query transformations: text-to-text (T2T), text-to-image (T2I),
and image-to-text (I2T) transformations.}
\label{exp:comparison_original_query_generative_query}
\centering
\resizebox{\linewidth}{!}{
\begin{tabular}{l|ccc|ccc|cc|c}
    \toprule
                       Datasets                      & \multicolumn{3}{c|}{IACC.3} & \multicolumn{3}{c|}{V3C1} & \multicolumn{2}{c|}{V3C2} &  \\
                       \hline
                       Query sets               & tv16  & tv17  & tv18  & tv19 & tv20 & tv21  & tv22 & tv23 & Mean  \\
                       \hline
                       User query &  0.280 &	0.349 &	0.165 &	0.242	&0.352	&0.365 &	0.235&	0.295 & 0.285\\ 
                       \hline
                       
                    T2T query  & 0.282      &   0.343    & 0.159      &  0.240     & 0.346 & 0.374& 0.221& 0.297&  0.283 \\
                        T2I query & 0.207&	0.236	&0.128	&0.215	&0.256	&0.270	&0.216 & 0.232&	0.220\\
                    I2T query &   0.233 &	0.328&	\textbf{0.185}	&0.214&	0.304	&0.323&	0.221&	0.282 &0.261\\

                    \hline
                                        User+T2T queries  &\textbf{0.285}	&0.357	&0.165	&0.244	&\textbf{0.361}	&0.372	&0.233	&0.301	&0.290\\

                    User+T2I queries    &0.274	&0.350	&0.171	&0.259	&0.352	&0.374	&0.250	&0.301	&0.291\\  
                    User+I2T queries     &0.277	&0.356	&0.182	&0.252	&0.347	&0.357	&0.252	&0.305	&0.291\\
                    \hline
                    User+T2T+T2I queries & 0.281 &0.356	&0.172	&0.260	&0.360	&0.382	&0.251	&0.305	&0.296\\

                    User+T2T+I2T queries    & 0.282 &0.360  &0.175  &0.255  &0.355 &0.370&	0.249&	0.308&	0.294\\  
                    User+T2I+I2T queries    &0.275	&0.351	&0.180	&0.260	&0.348	&0.365	&0.265	&0.310	&0.294\\

                     \hline
                   User+T2I+I2T+I2T queries       & 0.281	&\textbf{0.366}&0.180	&\textbf{0.263}	&0.359	&\textbf{0.377}	&\textbf{0.258}	&\textbf{0.311} &\textbf{0.299} \\  

  \bottomrule
\end{tabular}
}
\end{table}

\begin{table}[]
\caption{AVS performance comparison on logical, OOV and spatial queries between user and generated queries.}
\label{exp:comparison_original_query_generative_query_on_specificType}
\centering
\resizebox{\linewidth}{!}{

\begin{tabular}{l|c|c|c|c|c|c}
    \toprule

                    & Negation query & "AND" query & "OR" query & OOV query & Spatial query & Overall \\ \hline
User query & 0.082                   & 0.278                & 0.245               & 0.154              & 0.180                  & 0.188            \\ \hline
T2T query  & 0.086                   & \textbf{0.292}       & 0.240               & 0.166              & 0.189                  & 0.195            \\
T2I query  & 0.074                   & 0.217                & 0.198               & 0.145              & 0.177                  & 0.162            \\
I2T query  & \textbf{0.149}          & 0.253                &\textbf{ 0.246}      & \textbf{0.176}     & \textbf{0.228}         & \textbf{0.210}  \\
\bottomrule
\end{tabular}
}
\end{table}



Despite the bad performance of T2I when using it alone, its combination with user query (i.e., User+T2I queries) obtains a retrieval rate as high as other combinations, as shown in Table \ref{exp:comparison_original_query_generative_query}. Fusing all the rank lists retrieved by all paraphrased queries and the user query (i.e., Eq. (\ref{eq:ensemble}) obtains the best result. The randomization test~\cite{randomization_test} shows all seven ensembles have significant performance differences with the user query at $p$-value $\leq 0.01$. We further analyze how the rank changes before and after the ensemble. 
In general, the ensembles with paraphrased queries have improved the ranks of many user queries. Specifically, 48.03\%, 51.4\%, 52.61\%, and 55.69\% of true-positive videos receive higher ranks after the ensembles with the T2I, I2T, T2T, and all valid queries, respectively. Additionally, T2I, I2T, and T2T queries have different ways of improving the ranks of true positives in the ensembles. Specifically, the ensemble of the user and T2I queries has the lowest number of true positives getting evaluated. However, it has the largest magnitude of the change with a mean change of 139.00 versus 106.28 of user + T2T queries. Specifically, it has a larger number of true positives that have rank improvements by more than 500 ranks, which elevate true positives at the bottom section of the rank list to higher positions at the cost of some true positives having serious rank drops. In contrast, the ensembles with I2T and T2T queries (i.e., User + I2T queries and User + T2T queries) have more true positives with rank elevation but with smaller mean rank changes (i.e., 119.94 and 79.78, respectively). Moreover, it is worth mentioning that the ensembles with I2T and T2T queries bring around 7\% and 5\% of newly found true positives, which are out of the rank list of the user query, respectively. 

We also explore whether the query transformations introduce diversity to the retrieval results. Table \ref{exp:diversity_comparison} compares the diversity of the retrieved correct video sets across different combinations of ranked lists. Diversity is measured by computing the mean pairwise cosine distance of the IITV embedding features on a sub-sampled set. The sub-sampling is performed by randomly selecting an equal number of videos from different ranked lists. The correct video sets retrieved using T2I and I2T queries exhibit competitive diversity compared to the user query, and all incorporations of the transformations into the user query lead to improved diversity. The highest diversity is observed when all query transformations are applied simultaneously (User + T2T + I2T + T2I queries), achieving a significantly higher mean diversity score than the baseline (0.558 vs. 0.539). 

\begin{table}[]
\caption{Diversity comparison between different query transformations.}
\label{exp:diversity_comparison}
\centering
\resizebox{\linewidth}{!}{
\begin{tabular}{l|ccc|ccc|cc|c}
    \toprule
                       Datasets                      & \multicolumn{3}{c|}{IACC.3} & \multicolumn{3}{c|}{V3C1} & \multicolumn{2}{c|}{V3C2} &  \\
                       \hline
                       Query sets               & tv16  & tv17  & tv18  & tv19 & tv20 & tv21  & tv22 & tv23 & Mean  \\
                       \hline
User query                        & 0.541          & 0.548          & 0.616          & 0.538          & 0.526          & 0.529          & 0.501          & 0.510          & 0.539          \\ \hline
T2T query                         & 0.519          & 0.527          & 0.568          & 0.518          & 0.509          & 0.523          & 0.487          & 0.495          & 0.518          \\
T2I query                         & 0.542          & 0.544          & 0.619          & 0.531          & 0.521          & 0.530          & 0.499          & 0.508          & 0.537          \\
I2T query                         & 0.538          & 0.548          & 0.604          & 0.525          & 0.513          & 0.527          & 0.493          & 0.505          & 0.532          \\  \hline
User    + T2T query               & 0.550          & 0.562          & 0.616          & 0.549          & 0.539          & 0.542          & 0.515          & 0.522          & 0.549          \\
User    + T2I query               & 0.549          & 0.555          & 0.623          & 0.546          & 0.532          & 0.539          & 0.513          & 0.516          & 0.547          \\
User    + I2T query               & 0.546          & 0.556          & 0.628          & 0.549          & 0.532          & 0.537          & 0.518          & 0.521          & 0.548          \\ \hline
User    + T2T +  T2I query        & \textbf{0.556} & 0.558          & 0.627          & 0.558          & \textbf{0.546} & 0.543          & 0.519          & 0.527          & 0.554          \\
User    + T2T +  I2T query        & \textbf{0.556} & 0.565          & 0.625          & 0.556          & 0.540          & \textbf{0.552} & 0.523          & 0.527          & 0.556          \\
User    + I2T + T2I query         & 0.549          & \textbf{0.567} & 0.628          & 0.552          & 0.538          & 0.545          & 0.521          & 0.524          & 0.553          \\ \hline
User    + T2T + T2I + I2T queries & \textbf{0.556} & 0.564          & \textbf{0.632} & \textbf{0.561} & 0.543          & 0.551          & \textbf{0.527} & \textbf{0.528} & \textbf{0.558} \\
 
  \bottomrule
\end{tabular}
}
\end{table}


\subsubsection{Sensitivity of the Number of Valid Paraphrased Sub-queries in Video Search}
\label{exp_num_query}
We also explore how the number of valid queries affects the video retrieval performance. The experiment contrasts the AVS performance of using 1, 5, 15, 50, and 100 valid visual queries to retrieve videos. We found that using 15 valid visual queries significantly outperforms those less than 15. However, as the number of valid T2I queries increases to 50 or 100, the performances are not consistently better than that of 15 queries across all query sets. Using 50 valid T2I queries has no significant difference from using 15 valid T2I queries on retrieval performance. A similar trend is also observed in the I2T-to-video search. There is no significant difference between using 100 valid captioning queries and using 50 (or 25) valid ones. 

\section{Conclusion and Future Work}
\label{conclu}
This study has explored paraphrasing techniques to enhance the understanding of user queries. Empirical evaluations on AVS and Textual KIS tasks using TRECVid datasets demonstrate that the proposed query paraphrasing framework is effective for both short ad-hoc video queries and long-form known-item search queries. The T2I and I2T queries effectively address the out-of-vocabulary (OOV) problem and transform complex queries into simpler representations, thereby alleviating the query reasoning burden for the search engine. Additionally, the proposed T2I, I2T, and T2T queries complement the original user query by enhancing search intent comprehension and improving retrieval diversity, with T2I queries exhibiting the most significant divergence from the original query in identifying true positives. Furthermore, the proposed QA verification mechanism has been shown to effectively enhance retrieval performance by filtering out noisy queries introduced during the generation process.

However, this study also reveals certain limitations. First, the current T2I query approach has inherent challenges in capturing motion dynamics, particularly when multiple motions are involved. Additionally, the current image-to-video retrieval method is suboptimal, as it lacks query-specific attention to visual content and may overlook fine-grained attributes, such as color information. Addressing these limitations in future work will involve developing more advanced paraphrasing techniques and enhanced verification strategies to improve retrieval accuracy.

\bibliographystyle{IEEEtran}
\bibliography{IEEEabrv,ref}

\begin{thebibliography}{10}
\providecommand{\url}[1]{#1}
\csname url@samestyle\endcsname
\providecommand{\newblock}{\relax}
\providecommand{\bibinfo}[2]{#2}
\providecommand{\BIBentrySTDinterwordspacing}{\spaceskip=0pt\relax}
\providecommand{\BIBentryALTinterwordstretchfactor}{4}
\providecommand{\BIBentryALTinterwordspacing}{\spaceskip=\fontdimen2\font plus
\BIBentryALTinterwordstretchfactor\fontdimen3\font minus \fontdimen4\font\relax}
\providecommand{\BIBforeignlanguage}[2]{{%
\expandafter\ifx\csname l@#1\endcsname\relax
\typeout{** WARNING: IEEEtran.bst: No hyphenation pattern has been}%
\typeout{** loaded for the language `#1'. Using the pattern for}%
\typeout{** the default language instead.}%
\else
\language=\csname l@#1\endcsname
\fi
#2}}
\providecommand{\BIBdecl}{\relax}
\BIBdecl

\bibitem{Waseda2016}
K.~Ueki, K.~Kikuchi, S.~Saito, and T.~Kobayashi, ``{W}aseda at {TRECV}id 2016: Ad-hoc video search,'' in \emph{Proceedings of the {TRECV}id 2016 Workshop}, 2016, pp. 1--5.

\bibitem{Jiang:semantic_gap}
L.~Jiang, S.-I. Yu, D.~Meng, T.~Mitamura, and A.~Hauptmann, ``Bridging the ultimate semantic gap: A semantic search engine for internet videos,'' in \emph{Proceedings of the ACM International Conference on Multimedia Retrieval}, 2015.

\bibitem{Lu:ICMR}
Y.-J. Lu, H.~Zhang, M.~de~Boer, and C.-W. Ngo, ``Event detection with zero example: Select the right and suppress the wrong concepts,'' in \emph{Proceedings of the ACM on International Conference on Multimedia Retrieval}, 2016, pp. 127--134.

\bibitem{Informedia2016}
J.~Liang, P.~Huang, L.~Jiang, Z.~Lan, J.~Chen, and A.~Hauptmann, ``{I}nformedia @ {TRECV}id 2016 {MED} and {AVS},'' in \emph{Proceedings of the {TRECV}id 2016 Workshop}, 2016.

\bibitem{w2vvpp}
X.~Li, C.~Xu, G.~Yang, Z.~Chen, and J.~Dong, ``{W2VV}++: Fully deep learning for ad-hoc video search,'' in \emph{Proceedings of the ACM International Conference on Multimedia}, 2019, pp. 1786--1794.

\bibitem{tpami21-dual-encoding}
J.~Dong, X.~Li, C.~Xu, X.~Yang, G.~Yang, X.~Wang, and M.~Wang, ``Dual encoding for video retrieval by text,'' \emph{IEEE Transactions on Pattern Analysis and Machine Intelligence}, pp. 1--17, 2021.

\bibitem{shu_TMM2024MAC}
F.~Shu, B.~Chen, Y.~Liao, J.~Wang, and S.~Liu, ``Mac: Masked contrastive pre-training for efficient video-text retrieval,'' \emph{IEEE Transactions on Multimedia}, vol.~26, p. 9962–9972, Jan. 2024.

\bibitem{zhangTextVideoRetrievalGLSCL2025}
H.~Zhang, P.~Zeng, L.~Gao, J.~Song, Y.~Duan, X.~Lyu, and H.~T. Shen, ``Text-{{Video Retrieval With Global-LocalSemantic Consistent Learning}},'' \emph{IEEE Transactions on Image Processing}, vol.~34, pp. 3463--3474, 2025.

\bibitem{dual_task}
J.~Wu and C.-W. Ngo, ``Interpretable embedding for ad-hoc video search,'' in \emph{Proceedings of the ACM International Conference on Multimedia}, 2020, pp. 3357--3366.

\bibitem{TRECVid2017}
G.~Awad, A.~Butt, J.~Fiscus, D.~Joy, A.~Delgado, W.~McClinton, M.~Michel, A.~Smeaton, Y.~Graham, W.~Kraaij, G.~Quenot, M.~Eskevich, R.~Ordelman, G.~Jones, and B.~Huet, ``{TRECV}id 2017: Evaluating ad-hoc and instance video search, events detection, video captioning and hyperlinking,'' in \emph{Proceedings of the {TRECV}id 2017 Workshop}, 2018.

\bibitem{Trecvid2018}
G.~Awad, A.~Gov, A.~Butt, K.~Curtis, Y.~Lee, y.~Gov, J.~Fiscus, D.~Joy, A.~Delgado, A.~Smeaton, Y.~Graham, W.~Kraaij, G.~Quenot, J.~Magalhaes, and S.~Blasi, ``{TRECV}id 2018: Benchmarking video activity detection, video captioning and matching, video storytelling linking and video search,'' in \emph{Proceedings of the {TRECV}id 2018 Workshop}, 2018.

\bibitem{WasedaMeiseiSoftbank2019}
K.~Ueki, T.~Hori, and T.~Kobayashi, ``{W}aseda {M}eisei {S}oftbank at {TRECV}id 2019: Ad-hoc video search,'' in \emph{Proceedings of the {TRECV}id 2019 Workshop}, 2019, pp. 1--7.

\bibitem{wasedaAVS2020}
K.~Ueki, R.~Mutou, T.~Hori, Y.~Kim, and Y.~Suzuki, ``{W}aseda {M}eisei {S}oftbank at {TRECV}id 2020: Ad-hoc video search,'' in \emph{Proceedings of the {TRECV}id 2020 Workshop}, 2020, pp. 1--7.

\bibitem{Waseda2018}
K.~Ueki, Y.~Nakagome, K.~Hirakawa, K.~Kikuchi, Y.~Hayashi, T.~Ogawa, and T.~Kobayashi, ``{W}aseda {M}eisei at {TRECV}id 2018: Ad-hoc video search,'' in \emph{Proceedings of the {TRECV}id 2018 Workshop}, 2018, pp. 1--7.

\bibitem{improvedITV}
J.~Wu, C.~wah Ngo, and W.-K. Chan, ``Improving interpretable embeddings for ad-hoc video search with generative captions and multi-word concept bank,'' in \emph{Proceedings of the ACM on International Conference on Multimedia Retrieval}, 2024, pp. 1--10.

\bibitem{Informedia2018}
P.-Y. Huang, J.~Liang, Vaibhav, X.~Chang, and A.~Hauptmann, ``{I}nformedia @ {TRECV}id 2018: Ad-hoc video search with discrete and continuous representations,'' in \emph{Proceedings of the {TRECV}id 2018 Workshop}, 2018, pp. 1--10.

\bibitem{Trecvid2019}
G.~Awad, A.~Butt, K.~Curtis, Y.~Lee, J.~Fiscus, G.~Afzal, A.~Delgado, Z.~Jesse, E.~Godard, L.~Diduch, A.~F. Smeaton, Y.~Graham, W.~Kraaij, and G.~Quenot, ``{TRECV}id 2019: An evaluation campaign to benchmark video activity detection, video captioning and matching, and video search and retrieval,'' in \emph{Proceedings of the {TRECV}id 2019 Workshop}, 2019.

\bibitem{trecvid2023}
G.~Awad, K.~Curtis, A.~A. Butt, J.~Fiscus, A.~Godil, Y.~Lee, A.~Delgado, E.~Godard, L.~Diduch, Y.~Graham, , and G.~Quénot, ``Trecvid 2023 - a series of evaluation tracks in video understanding,'' in \emph{Proceedings of TRECVID 2023}.\hskip 1em plus 0.5em minus 0.4em\relax NIST, USA, 2023.

\bibitem{Understand_Negation_rumcc_MM}
Z.~Wang, A.~Chen, F.~Hu, and X.~Li, ``Learn to understand negation in video retrieval,'' in \emph{Proceedings of the 30th ACM International Conference on Multimedia}, 2022, p. 434–443.

\bibitem{LaCLIP}
L.~Fan, D.~Krishnan, P.~Isola, D.~Katabi, and Y.~Tian, ``Improving clip training with language rewrites,'' in \emph{NeurIPS}, 2023.

\bibitem{robustness_data_augumentation}
S.~Ramshetty, G.~Verma, and S.~Kumar, ``Cross-modal attribute insertions for assessing the robustness of vision-and-language learning,'' in \emph{Proceedings of the 61st Annual Meeting of the Association for Computational Linguistics (Volume 1: Long Papers)}, Jul. 2023, pp. 15\,974--15\,990.

\bibitem{Waseda2023}
K.~Ueki, Y.~Suzuki, H.~Takushima, H.~Sato, T.~Takada, H.~Okamoto, H.~Tanoue, T.~Hori, and A.~M. Kumar, ``{W}aseda meisei softbank at trecvid 2023,'' in \emph{Proceedings of the {TRECV}id 2023 Workshop}, 2023, pp. 1--8.

\bibitem{rombach2021stablediffusion}
R.~Rombach, A.~Blattmann, D.~Lorenz, P.~Esser, and B.~Ommer, ``High-resolution image synthesis with latent diffusion models,'' in \emph{2022 IEEE/CVF Conference on Computer Vision and Pattern Recognition (CVPR)}, jun 2022, pp. 10\,674--10\,685.

\bibitem{BLIP2}
J.~Li, D.~Li, S.~Savarese, and S.~Hoi, ``Blip-2: Bootstrapping language-image pre-training with frozen image encoders and large language models,'' \emph{arXiv 2301.12597}, 2023.

\bibitem{Trecvid2016}
G.~Awad, F.~Jonathan, J.~David, M.~Martial, S.~Alan, K.~Wessel, Q.~Georges, E.~Maria, A.~Robin, O.~Roeland, J.~Gareth, H.~Benoit, and LarsonMartha, ``{TRECV}id 2016: Evaluating video search, video event detection, localization, and hyperlinking,'' in \emph{Proceedings of the {TRECV}id 2016 Workshop}, 2016, pp. 1--54.

\bibitem{V3C1}
F.~Berns, L.~Rossetto, K.~Schoeffmann, C.~Beecks, and G.~Awad, ``{V3C1} dataset: An evaluation of content characteristics,'' in \emph{Proceedings of the International Conference on Multimedia Retrieval}, 2019, pp. 334--338.

\bibitem{trecvid2006}
A.~F. Smeaton, P.~Over, and W.~Kraaij, ``Evaluation campaigns and {TRECV}id,'' in \emph{Proceedings of the ACM International Workshop on Multimedia Information Retrieval}, 2006, pp. 321--330.

\bibitem{Jiang2007VIREO374L}
Y.-G. Jiang, J.~Yang, C.-W. Ngo, and A.~G. Hauptmann, ``Representations of keypoint-based semantic concept detection: A comprehensive study,'' \emph{IEEE Transactions on Multimedia}, vol.~12, no.~1, pp. 42--53, 2010.

\bibitem{vse}
F.~Faghri, D.~J. Fleet, J.~R. Kiros, and S.~Fidlere, ``{VSE}++: Improving visual-semantic embeddings with hard negatives,'' in \emph{Proceedings of the British Machine Vision Conference}, 2018, pp. 1--13.

\bibitem{dualconding}
J.~Dong, X.~Li, C.~Xu, S.~Ji, Y.~He, G.~Yang, and X.~Wang, ``Dual encoding for zero-example video retrieval,'' in \emph{Proceedings of the IEEE Conference on Computer Vision and Pattern Recognition}, 2019, pp. 9346--9355.

\bibitem{LAFF}
F.~Hu, A.~Chen, Z.~Wang, F.~Zhou, J.~Dong, and X.~Li, ``Lightweight attentional feature fusion: A new baseline for text-to-video retrieval,'' in \emph{European Conference on Computer Vision}.\hskip 1em plus 0.5em minus 0.4em\relax Cham: Springer Nature Switzerland, 2022, pp. 444--461.

\bibitem{wu2023ITV}
J.~Wu, C.-W. Ngo, W.-K. Chan, and Z.~Hou, ``(un)likelihood training for interpretable embedding,'' in \emph{ACM Transactions on Information Systems}, 2023.

\bibitem{RIVRL}
J.~Dong, Y.~Wang, X.~Chen, X.~Qu, X.~Li, Y.~He, and X.~Wang, ``Reading-strategy inspired visual representation learning for text-to-video retrieval,'' \emph{IEEE Transactions on Circuits and Systems for Video Technology}, vol.~32, p. 5680–5694, 2022.

\bibitem{liMultiModalInductiveFramework2024a}
Q.~Li, Y.~Zhou, C.~Ji, F.~Lu, J.~Gong, S.~Wang, and J.~Li, ``Multi-{{Modal Inductive Framework}} for {{Text-Video Retrieval}},'' in \emph{Proceedings of the 32nd {{ACM International Conference}} on {{Multimedia}}}.\hskip 1em plus 0.5em minus 0.4em\relax ACM, Oct. 2024, pp. 2389--2398.

\bibitem{hgr}
S.~{Chen}, Y.~{Zhao}, Q.~{Jin}, and Q.~{Wu}, ``Fine-grained video-text retrieval with hierarchical graph reasoning,'' in \emph{Proceedings of the IEEE/CVF Conference on Computer Vision and Pattern Recognition}, 2020, pp. 10\,635--10\,644.

\bibitem{queryParaphsingDocument}
I.~Zukerman and B.~Raskutti, ``Lexical query paraphrasing for document retrieval,'' in \emph{Proceedings of the 19th International Conference on Computational Linguistics}, 2002, p. 1–7.

\bibitem{queryExpansionObjectRetrieval}
O.~Chum, J.~Philbin, J.~Sivic, M.~Isard, and A.~Zisserman, ``Total recall: Automatic query expansion with a generative feature model for object retrieval,'' in \emph{2007 IEEE 11th International Conference on Computer Vision}, 2007, pp. 1--8.

\bibitem{VBS2018}
J.~Loko\v{c}, G.~Koval\v{c}\'{\i}k, B.~M\"{u}nzer, K.~Sch\"{o}ffmann, W.~Bailer, R.~Gasser, S.~Vrochidis, P.~A. Nguyen, S.~Rujikietgumjorn, and K.~U. Barthel, ``Interactive search or sequential browsing? a detailed analysis of the video browser showdown 2018,'' \emph{ACM Trans. Multimedia Comput. Commun. Appl.}, vol.~15, no.~1, 2019.

\bibitem{wang2023querydoc}
L.~Wang, N.~Yang, and F.~Wei, ``Query2doc: Query expansion with large language models,'' in \emph{The 2023 Conference on Empirical Methods in Natural Language Processing}, 2023.

\bibitem{OpenAI2023ChatGPT}
{OpenAI}, ``Chatgpt,'' \url{https://openai.com}, 2023.

\bibitem{touvron2023llama}
H.~Touvron, T.~Lavril, G.~Izacard, X.~Martinet, M.-A. Lachaux, T.~Lacroix, B.~Rozière, N.~Goyal, E.~Hambro, F.~Azhar, A.~Rodriguez, A.~Joulin, E.~Grave, and G.~Lample, ``Llama: Open and efficient foundation language models,'' 2023.

\bibitem{CLIP}
A.~Radford, J.~W. Kim, C.~Hallacy, A.~Ramesh, G.~Goh, S.~Agarwal, G.~Sastry, A.~Askell, P.~Mishkin, J.~Clark, G.~Krueger, and I.~Sutskever, ``Learning transferable visual models from natural language supervision,'' in \emph{Proceedings of the 38th International Conference on Machine Learning, {ICML}}, 2021, pp. 8748--8763.

\bibitem{VIREO_VBS2024}
Z.~Ma, J.~Wu, and C.~W. Ngo, ``Leveraging llms and generative models for interactive known-item video search,'' in \emph{MultiMedia Modeling}, 2024, pp. 380--386.

\bibitem{genSceneImageForVideoSum}
R.~Yanagi, R.~Togo, T.~Ogawa, and M.~Haseyama, ``Scene retrieval for video summarization based on text-to-image gan,'' in \emph{2019 IEEE International Conference on Image Processing (ICIP)}, 2019, pp. 1825--1829.

\bibitem{askAndConfirm_ICCV}
G.~Cai, J.~Zhang, X.~Jiang, Y.~Gong, L.~He, F.~Yu, P.~Peng, X.~Guo, F.~Huang, and X.~Sun, ``Ask\&confirm: Active detail enriching for cross-modal retrieval with partial query,'' in \emph{Proceedings of the IEEE/CVF International Conference on Computer Vision}, 2021, pp. 1835--1844.

\bibitem{QAforRetrieval_MM}
A.~Madasu, J.~Oliva, and G.~Bertasius, ``Learning to retrieve videos by asking questions,'' in \emph{Proceedings of the 30th ACM International Conference on Multimedia}, ser. MM '22, 2022, p. 356–365.

\bibitem{liang2023simplebaselineQA}
K.~Liang and S.~Albanie, ``Simple baselines for interactive video retrieval with questions and answers,'' in \emph{Proceedings of the IEEE/CVF International Conference on Computer Vision}, 2023, pp. 11\,091--11\,101.

\bibitem{openai2023gpt4}
OpenAI, ``{GPT-4} technical report,'' \emph{CoRR}, vol. abs/2303.08774, 2023.

\bibitem{Waseda_Meisei2017}
K.~Ueki, K.~Hirakawa, K.~Kikuchi, T.~Ogawa, and T.~Kobayashi, ``{W}aseda {M}eisei at {TRECV}id 2017: Ad-hoc video search,'' in \emph{Proceedings of the {TRECV}id 2017 Workshop}, 2017, pp. 1--8.

\bibitem{videostory}
A.~Habibian, T.~Mensink, and C.~G.~M. Snoek, ``{V}ideo{S}tory: A new multimedia embedding for few-example recognition and translation of events,'' in \emph{Proceedings of the ACM Conference on Multimedia}, 2014, pp. 17--26.

\bibitem{w2vv}
J.~Dong, X.~Li, and C.~G.~M. Snoek, ``Predicting visual features from text for image and video caption retrieval,'' \emph{IEEE Transactions on Multimedia}, vol.~20, no.~12, pp. 3377--3388, 2018.

\bibitem{W2VVpp_Case_Study}
J.~Loko\'{c}, T.~Sou\'{c}ek, P.~Vesel\'{y}, F.~Mejzl\'{\i}k, J.~Ji, C.~Xu, and X.~Li, ``A {W2VV}++ case study with automated and interactive text-to-video retrieval,'' in \emph{Proceedings of the 28th ACM International Conference on Multimedia}, 2020, pp. 2553–--2561.

\bibitem{dual_coding_improved}
D.~Galanopoulos and V.~Mezaris, ``Attention mechanisms, signal encodings and fusion strategies for improved ad-hoc video search with dual encoding networks,'' in \emph{Proceedings of the 2020 International Conference on Multimedia Retrieval}, 2020, pp. 336–--340.

\bibitem{tmm2021-sea}
X.~Li, F.~Zhou, C.~Xu, J.~Ji, and G.~Yang, ``{SEA}: Sentence encoder assembly for video retrieval by textual queries,'' \emph{IEEE Transactions on Multimedia}, pp. 4351--4362, 2021.

\bibitem{Luo2021CLIP4Clip}
H.~Luo, L.~Ji, M.~Zhong, Y.~Chen, W.~Lei, N.~Duan, and T.~Li, ``{CLIP4Clip}: An empirical study of clip for end to end video clip retrieval,'' \emph{arXiv preprint arXiv:2104.08860}, pp. 1--14, 2021.

\bibitem{zhu2024languagebind}
B.~Zhu, B.~Lin, M.~Ning, Y.~Yan, J.~Cui, W.~HongFa, Y.~Pang, W.~Jiang, J.~Zhang, Z.~Li, C.~W. Zhang, Z.~Li, W.~Liu, and L.~Yuan, ``Languagebind: Extending video-language pretraining to n-modality by language-based semantic alignment,'' in \emph{The Twelfth International Conference on Learning Representations}, 2024.

\bibitem{randomization_test}
M.~D. Smucker, J.~Allan, and B.~Carterette, ``A comparison of statistical significance tests for information retrieval evaluation,'' in \emph{Proceedings of the ACM Conference on Conference on Information and Knowledge Management}, 2007, pp. 623--632.

\end{thebibliography}

\newpage

 




\vfill

\end{document}